\documentclass[journal=jpclcd, layout=traditional]{achemso}%

\usepackage[T1]{fontenc} 

\usepackage{amsmath,color}
\usepackage{amssymb}
\usepackage{amsfonts}
\usepackage{amsthm}
\usepackage{mathtools}
\usepackage{dcolumn}
\usepackage{bm}

\usepackage{rotating}

\DeclarePairedDelimiter\ket{\lvert}{\rangle}
\DeclarePairedDelimiterX\braket[2]{\langle}{\rangle}{#1 \delimsize\vert #2}

\newcommand{\vJ}{\mathbf{J}}
\newcommand{\vP}{{\boldsymbol{\mathcal{P}}}}
\newcommand{\vE}{\mathbf{E}}
\newcommand{\vB}{\mathbf{B}}
\newcommand{\vD}{\mathbf{D}}
\newcommand{\ve}{\mathbf{e}}
\newcommand{\vnabla}{\boldsymbol{\nabla}}
\newcommand{\vdelta}{\boldsymbol{\delta}}
\newcommand{\vmu}{\boldsymbol{\mu}}

\newcommand{\vR}{\mathbf{r}}
\newcommand{\vk}{\mathbf{k}}

\newcommand{\vxi}{\boldsymbol{\xi}}
\newcommand{\kFGR}{k_{\text{FGR}}}
\newcommand{\kEh}{k_{\text{Eh}}}

\newcommand{\hH}{\hat{H}}
\newcommand{\hD}{\hat{\mathbf{D}}}
\newcommand{\hE}{\hat{\mathbf{E}}}
\newcommand{\hB}{\hat{\mathbf{B}}}
\newcommand{\hA}{\hat{\mathbf{A}}}
\newcommand{\hrho}{\hat{\rho}}
\newcommand{\hdelta}{\hat{\delta}_{\mathbf{E}}}
\newcommand{\hV}{\hat{V}}

\newcommand{\hDperp}{\hat{\mathbf{D}}_{\perp}}
\newcommand{\hEperp}{\hat{\mathbf{E}}_{\perp}}
\newcommand{\hPperp}{\hat{\boldsymbol{\mathcal{P}}}_{\perp}}
\newcommand{\hP}{\hat{\boldsymbol{\mathcal{P}}}}

\newcommand{\tr}[1]{\text{Tr}\left(#1\right)}
\newcommand{\avg}[1]{\left\langle #1\right\rangle}

\title{A Necessary Trade-off for Semiclassical Electrodynamics:  Accurate Short-Range Coulomb Interactions versus the Enforcement of Causality?}%

\author{Tao E. Li}%
\email{taoli@sas.upenn.edu}
\affiliation{Department of Chemistry, University of Pennsylvania, Philadelphia, Pennsylvania 19104, USA}
\author{Hsing-Ta Chen}
\affiliation{Department of Chemistry, University of Pennsylvania, Philadelphia, Pennsylvania 19104, USA}
\author{Abraham Nitzan}
\affiliation{Department of Chemistry, University of Pennsylvania, Philadelphia, Pennsylvania 19104, USA}
\author{Maxim Sukharev}
\affiliation{Department of Physics, Arizona State University, Tempe, Arizona 85287, USA}
\alsoaffiliation{College of Integrative Sciences and Arts, Arizona State University, Mesa, AZ 85212, USA}
\author{Joseph E. Subotnik}
\email{subotnik@sas.upenn.edu}
\affiliation{Department of Chemistry, University of Pennsylvania, Philadelphia, Pennsylvania 19104, USA}

\begin{document}

\begin{tocentry}
	\includegraphics[width=5.1cm]{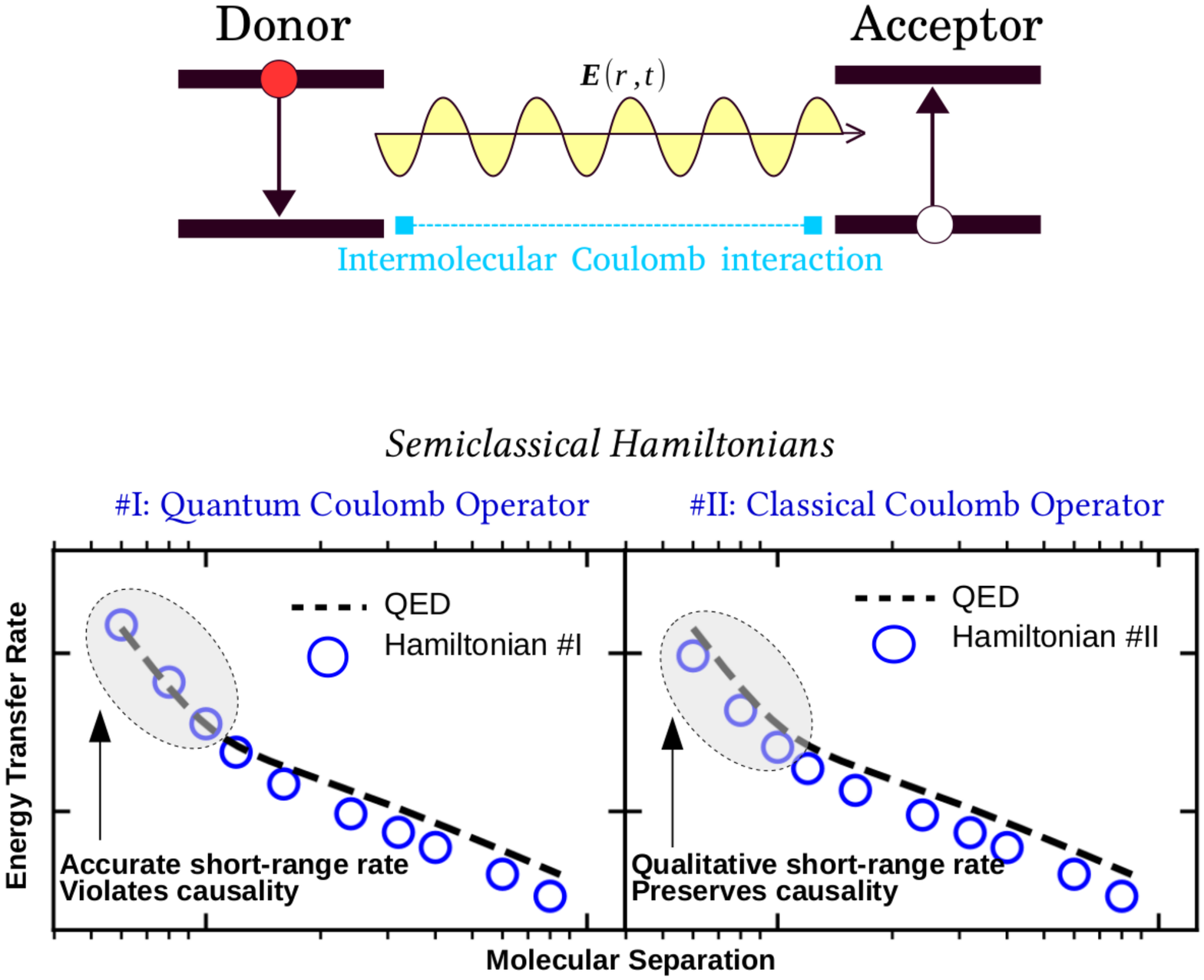}
\end{tocentry}

\begin{abstract}
	We investigate two key representative semiclassical approaches for propagating resonant energy transfer (RET) between a pair of electronic two-level systems (donor and acceptor) with coupled Maxwell-Liouville equations. On the one  hand, when the electromagnetic (EM) field is treated classically and Coulomb interactions are treated quantum-mechanically,  we find that a quantum-classical mismatch leads to a violation of causality, i.e., the acceptor can be excited before the retarded EM field arrives. 
	On the other hand, if we invoke a classical intermolecular Coulomb operator, we find that the energy transfer in the near field loses quantitative accuracy compared with  F\"orster theory, even though causality is strictly obeyed.  Thus, our work raises a fundamental paradox when choosing  a semiclassical electrodynamics algorithm. Namely, which is more important: Accurate short range interactions or long range causality?
	Apparently, one cannot have one's cake and eat it too.
\end{abstract}

Light-matter interactions are an essential research area in physics, chemistry and engineering. A host of recent experiments  encountering strong light-matter interactions\cite{Torma2015, Sukharev2017, Vasa2018, Khitrova2006, Gibbs2011, Lodahl2015, Ribeiro2018} have demonstrated that the optical response of matter does not always follow response theory, and that we cannot always treat the electromagnetic (EM) field as a perturbation\cite{weisskopf1930calculation, meystre1992theoretical, Meystre2007, thakkar2015quantum}.  In order to model such experiments,  an optimal approach should consider both the light and matter degrees of freedom on the same footing. 

For a non-perturbative model of electrodynamics in terms of molecular properties, the usual approach is to perform  a Power-Zienau-Woolley (PZW) transformation\cite{Power1983-1, Mukamel1999}, so that the full quantum electrodynamics (QED) Hamiltonian reads as follows,
\begin{equation}\label{eq:PZW_H}
\begin{aligned}
\hH &= \hH_s + \frac{1}{2}\int d\vR\ \left[ \frac{|\hDperp(\vR)|^2}{\epsilon_0} + \frac{|\hB(\vR)|^2}{\mu_0} \right] \\
&- \int d\vR\ \frac{\hDperp(\vR)}{\epsilon_0}\hPperp(\vR) + \frac{1}{2\epsilon_0}\int d\vR \ |\hPperp(\vR)|^2
\end{aligned}
\end{equation} 
Here, we ignore the magnetic and diamagnetic interactions for the quantum subsystem. $\hDperp$  and $\hB$ are  the displacement and magnetic field operators,  $\hH_s$ is the Hamiltonian for the quantum subsystem, and $\hPperp$ is the transverse polarization operator of the quantum (molecular) subsystem that couples to the EM field.\cite{footnote3} Note that the transverse component of $\hP$ satisfies  $\vnabla \cdot \hPperp = 0$ and the longitudinal component of $\hP$ satisfies $\vnabla \times \hP_{\parallel} = 0$. $\hDperp = \epsilon_0\hEperp + \hP_{\perp}$  and $\hB = \vnabla\times \hA$, where $\hA$ is the vector potential. The canonical commutator relationship is $\left[\hDperp(\vR), \hA(\vR')\right] = i\hbar\vdelta_{\perp}(\vR - \vR')$, where $\vdelta_{\perp}$ is the transverse $\delta$-function. Formally, the regularized transverse $\delta$-function can be written as $\delta_{\perp ij}(\vR) = \frac{2}{3}\delta_{ij}\delta(\vR) + \frac{\eta(\vR)}{4\pi r^3}\left(\frac{3r_i r_j}{r^2} - \delta_{ij}\right)$, where $i, j = x, y, z$ and $\eta(\vR)$ is $0$ at $\vR = \mathbf{0}$ to suppress the divergence (but $\eta(\vR)$ equals $1$ elsewhere).\cite{Cohen-Tannoudji1997} Note that for a neutral system, the displacement field is exclusively transverse, (i.e., $\hD_{\parallel} = 0$), so that we can write $\hD$ or $\hDperp$ interchangeably. Although not discussed often, we note that Eq. (\ref{eq:PZW_H}) should formally include the self-interaction of all charges (which is infinitely large unless one introduces a cutoff); see Eqs.  (I.B.36) and (IV.C.38) in Ref. \citenum{Cohen-Tannoudji1997}.

At this point, let us consider  a system containing $N$ separable and neutral molecules. Here, one can write:
\begin{equation}\label{eq:many_particle_expression}
\begin{aligned}
\hH_s &= \sum_{n=1}^{N}\hH_s^{(n)} + \sum_{n < l}\hV_{Coul}^{(nl)} \\
\hPperp &= \sum_{n=1}^{N}\hPperp^{(n)}
\end{aligned}
\end{equation}
where  the intermolecular Coulomb interactions $\hV_{Coul}^{(nl)} $ are (for $n \ne l$) \cite{Cohen-Tannoudji1997}
\begin{equation}\label{eq:Vdd_exp_1}
\hV_{Coul}^{(nl)} = \frac{1}{\epsilon_0}\int d\vR\  \hP^{(n)}_{\parallel}(\vR)\cdot \hP^{(l)}_{\parallel}(\vR)
\end{equation}
In Eq. (\ref{eq:Vdd_exp_1}), the intermolecular Coulomb operator is defined as the inner product of the longitudinal  polarization operators for
the molecules $n$ and $l$. 
When the molecular size is much less than the intermolecular separation, one can make the point-dipole approximation, i.e., $\hP^{(n)}(\vR) = \hat{\vmu}^{(n)}\vdelta(\vR - \vR^{(n)})$. The longitudinal  polarization operator is then $\hP^{(n)}_{\parallel}(\vR) = \hat{\vmu}^{(n)}\vdelta_{\parallel}(\vR - \vR^{(n)}) = \sum\limits_{i, j}\ve_i \left[-\frac{1}{3}\delta_{ij}\delta(\vR - \vR^{(n)}) - \frac{\eta(\vR - \vR^{(n)})}{4\pi |\vR - \vR^{(n)}|^3}\left(\frac{3\left(\vR_i - \vR^{(n)}_i\right)\left(\vR_j - \vR^{(n)}_j\right)}{|\vR - \vR^{(n)}|^2} - \delta_{ij}\right)\right]\hat{\mu}_j$. Therefore, Eq. (\ref{eq:Vdd_exp_1}) can be reduced to the well-known instantaneous dipole-dipole interaction Hamiltonian\cite{footnoteProve}: 
\begin{equation}\label{eq:Vdd_interaction_point}
\hV_{Coul}^{(nl)} = \frac{1}{4\pi \epsilon_0}\left(\frac{\hat{\vmu}^{(n)}\cdot \hat{\vmu}^{(l)}}{|\vR|^3} - \frac{3(\hat{\vmu}^{(n)}\cdot \hat{\vR})(\hat{\vmu}^{(l)} \cdot \hat{\vR})}{|\vR|^3}\right)
\end{equation}
Here, $\hat{\vmu}^{(n, l)}$  is the dipole moment operator of molecule $n$ or $l$ and $\vR$ ($\hat{\vR}$) is the vector (unit vector) along the direction of molecular separation. 

At this point, one can prove causality through the following argument.  Consider the case of two molecules well separated from each other
(so that $\int d\vR \ \hP^{(n)}\cdot \hP^{(l)}= 0$). Then, 
if we substitute Eqs. (\ref{eq:many_particle_expression}) and (\ref{eq:Vdd_exp_1}) into Eq. (\ref{eq:PZW_H}), we find that
all instantaneous interactions between molecular pairs  vanish by cancellation:
\begin{equation}\label{eq:PZW_H_2}
\begin{aligned}
\hH = &\sum_{n=1}^{N}\hH_s^{(n)} + \frac{1}{2}\int d\vR\ \left[ \frac{|\hDperp(\vR)|^2}{\epsilon_0} + \frac{|\hB(\vR)|^2}{\mu_0} \right] - \sum_{n=1}^{N}\int d\vR\ \frac{\hDperp(\vR)}{\epsilon_0}\hPperp^{(n)}(\vR) + \\ &\sum_{n=1}^{N}\frac{1}{2\epsilon_0}\int d\vR \ |\hPperp^{(n)}(\vR)|^2
\end{aligned}
\end{equation}
where we have used the identity
\begin{equation}\label{eq:PZW_V_inter_cancel}
\begin{aligned}
&\hV_{Coul}^{(nl)}  + \frac{1}{\epsilon_0}\int d\vR \ \hP_{\perp}^{(n)}\cdot\hP_{\perp}^{(l)} \\
= &\frac{1}{\epsilon_0}\int d\vR \ \hP_{\parallel}^{(n)}\cdot\hP_{\parallel}^{(l)}  + \frac{1}{\epsilon_0}\int d\vR \ \hP_{\perp}^{(n)}\cdot\hP_{\perp}^{(l)}\\
	= &\frac{1}{\epsilon_0}\int d\vR \ \hP^{(n)}\cdot\hP^{(l)}\\
	= &0
\end{aligned}
\end{equation}
Thus, QED strictly satisfies causality: molecules interact solely through the retarded EM field. The Hamiltonians in Eqs. (\ref{eq:PZW_H}) and (\ref{eq:PZW_H_2}) are identical.

\paragraph{A semiclassical algorithm for QED: the lack of a unique approach.}
When dealing with realistically large systems, the many-body Hamiltonian in Eqs. (\ref{eq:PZW_H}) and (\ref{eq:PZW_H_2}) are 
almost impossible 
to propagate quantum-mechanically, and the only practical method is usually time-dependent perturbation theory with small light-matter interactions. To overcome this restriction, one promising approach is to use semiclassical electrodynamics, whereby one treats the EM field classically while treating the molecular subsystem quantum mechanically and there is no small parameter\cite{Masiello2005, Lopata2009, Sukharev2011, Puthumpally-Joseph2015, Li2018Spontaneous}. According to this approach, one evolves the coupled Schr\"odinger-Maxwell or Liouville-Maxwell equations:
\begin{subequations}\label{eq:Ehrenfest}
	\begin{align}
	\frac{d}{dt}\hat{\rho}(t) &= -\frac{i}{\hbar}\left[\hH_{sc}(t),\ \hat{\rho}(t)\right] \label{eq:Hdef}\\
	\frac{\partial}{\partial t}\vB(\vR, t) &= -\vnabla \times \vE(\vR, t) \\
	\frac{\partial}{\partial t}\vE(\vR, t) &= c^2 \vnabla \times \vB(\vR, t) - \frac{\vJ(\vR, t)}{\epsilon_0}  \label{eq:J0} \\
	\vJ(\vR, t) &= \frac{d}{dt}\tr{\hat{\rho}(t) \hP(\vR)} \label{eq:J}
	\end{align}
\end{subequations}
Here, $\hat{\rho}$, $\hH_{sc}$ and $\hP$ are (respectively) the density operator, the semiclassical Hamiltonian and the polarization operator for the quantum molecular subsystem.
For a subsystem containing $N$ molecules, the total density operator $\hat{\rho} $ is expressed as $\hat{\rho} = \hat{\rho}^{(1)}\otimes \hat{\rho}^{(2)}\otimes \cdots\otimes \hat{\rho}^{(N)}$. In Eq. (\ref{eq:J0}),  $c = 1/\sqrt{\mu_0\epsilon_0}$ and  $\vJ$ is the current density operator that connects the quantum molecular subsystem to the classical EM field.  
In Eq. (\ref{eq:J}),  $\vJ$ is defined by a mean-field approximation\cite{Prezhdo1997, Li2005}, and so the set Eqs. (\ref{eq:Ehrenfest}) can also be called ``Ehrenfest'' electrodynamics. As far as the notation below, it will be crucial to distinguish between the operator $\hP$ (with hat)  and the average $\vP = \tr{\hat{\rho}\hP}$ (no hat).

Note that Eq. (\ref{eq:J0}) can be separated into two different equations for the transverse and perpendicular components:
\begin{subequations}
	\begin{align}
	\frac{\partial}{\partial t}\vE_{\perp}(\vR, t) &= c^2 \vnabla \times \vB(\vR, t) - \frac{\vJ_{\perp}(\vR, t)}{\epsilon_0}  \\
	\frac{\partial}{\partial t}\vE_{\parallel}(\vR, t) &= - \frac{\vJ_{\parallel}(\vR, t)}{\epsilon_0} 
	\end{align}
\end{subequations}
and the latter equation can be integrated so that:
\begin{equation}
\vE_{\parallel}(\vR, t) = -\frac{\vP_{\parallel}(\vR, t)}{\epsilon_0}
\end{equation}

\paragraph{Hamiltonian \#I.}
When defining the semiclassical, electronic Hamiltonian $\hH_{sc}$ in Eq. (\ref{eq:Hdef}), there is no unique prescription. In the supporting information, we provide a detailed approach for constructing two different semiclassical Hamiltonians starting from the PWZ Hamiltonian. Here, we present only the main results.

The first  Hamiltonian\cite{Mukamel1999} reads
\begin{equation}\label{eq:Hsc-Escheme}
\hH_{sc}^{I} = \sum_{n=1}^{N} \left[\hH_s^{(n)} - \int d\vR \vE_{\perp}(\vR, t)\cdot \hP^{(n)}(\vR)\right] + \sum_{n <  l}\hV_{Coul}^{(nl)}
\end{equation}
Henceforward, we will refer to Eq. (\ref{eq:Hsc-Escheme}) as Hamiltonian \#I. 

In Eq. (\ref{eq:Hsc-Escheme}),  there are two terms containing instantaneous interactions: the non-local transverse E-field ($\vE_{\perp}$) and the intermolecular Coulomb interactions ($\hV_{Coul}^{(nl)}$).  Just as for QED, one would normally expect that Eqs. (\ref{eq:Ehrenfest}-\ref{eq:Hsc-Escheme}) should preserve causality. 
This alleged cancellation should be obvious if we substitute in $\vE_{\perp} = \vE - \vE_{\parallel} = \vE + \frac{1}{\epsilon_0}\vP_{\parallel}$, 
so that we can rewrite Eq. (\ref{eq:Hsc-Escheme}) as:
\begin{equation}\label{eq:Hsc_E_simulate}
\begin{aligned}
\hH_{sc}^{I} &=  \sum_{n=1}^{N} \left[\hH_s^{(n)} - \int d\vR \left(\vE(\vR, t)  +\frac{1}{\epsilon_0} \vP_{\parallel}^{(n)}(\vR) \right)\cdot \hP^{(n)}(\vR)\right]  \\ &
    - \frac{1}{\epsilon_0}\sum_{n\neq l}\int d\vR\vP_{\parallel}^{(n)}(\vR) \cdot \hP^{(l)}(\vR) + \sum_{n <  l} \hV_{Coul}^{(nl)}  
\color{black} 
\end{aligned}
\end{equation}
Ideally,  the second line of Eq. (\ref{eq:Hsc_E_simulate}) should cancel (see Eq. (\ref{eq:PZW_V_inter_cancel})).  
However, note that in Eq. (\ref{eq:Hsc_E_simulate}), one of the $\vP$ terms is treated
classically while the Coulomb interactions are treated fully quantum-mechanically (see Eq. (\ref{eq:Vdd_exp_1})), and thus, there is no guarantee of cancellation or strict causality.
In fact,  below we will present numerical simulations showing that causality is not strictly enforced.
Thus, one may further ask: can we find a different semiclassical Hamiltonian that does preserve causality? Indeed, this is possible, which brings us to  Hamiltonian \#II.

\paragraph{Hamiltonian \#II.}
To preserve causality, one can make the  following approximation:  $\forall  n, l$,
\begin{equation}\label{eq:Vdd}
\begin{aligned}
\hV_{Coul}^{(nl)} & = \frac{1}{\epsilon_0}\int d\vR \vP^{(n)}_{\parallel}(\vR, t)\cdot \hP_{\parallel}^{(l)}(\vR) \\
&+ \frac{1}{\epsilon_0}\int d\vR \vP^{(l)}_{\parallel}(\vR, t)\cdot \hP^{(n)}_{\parallel}(\vR)
\end{aligned}
\end{equation}
Compared with the quantum form of $\hV_{Coul}^{(nl)} $ in Eq. (\ref{eq:Vdd_exp_1}), the physical meaning of Eq. (\ref{eq:Vdd}) is clear: the intermolecular Coulomb interactions
between molecules are effectively the classical polarization energies as felt by one molecule in the field of another and as expressed by the 
classical longitudinal  polarization fields ($\vP_{\parallel}^{(n)}$ and $\vP_{\parallel}^{(l)}$). 
If we substitute Eq. (\ref{eq:Vdd}) and $\vE_{\perp} = \frac{1}{\epsilon_0}(\vD - \vP_{\perp})$ into Eq. (\ref{eq:Hsc-Escheme}), after some straightforward algebra, we find that 
a new semiclassical Hamiltonian emerges
\begin{equation}\label{eq:Hsc-Dscheme}
\begin{aligned}
\hH_{sc}^{II} = \sum_{n=1}^{N} &\hH_s^{(n)} - \frac{1}{\epsilon_0}\int d\vR \vD(\vR, t)\cdot \hP^{(n)}(\vR) \\
&+ \frac{1}{\epsilon_0}\int d\vR \vP_{\perp}^{(n)}(\vR, t)\cdot \hP^{(n)}(\vR)
\end{aligned}
\end{equation}
In Eq. (\ref{eq:Hsc-Dscheme}), the intermolecular interactions are carried exclusively through the classical D-field, and thus  causality is strictly preserved. 
Henceforward, to distinguish  Eq. (\ref{eq:Hsc-Dscheme}) from
Eq. (\ref{eq:Hsc-Escheme}), we will refer to Eq. (\ref{eq:Hsc-Dscheme})  as Hamiltonian \#II. Note that, by substituting Eq. (\ref{eq:Vdd}) into Eq. (\ref{eq:Hsc_E_simulate}), Eq. (\ref{eq:Hsc-Dscheme}) is equivalent to
\begin{equation}\label{eq:Hsc_D_simulate}
\begin{aligned}
\hH_{sc}^{II} &=  \sum_{n=1}^{N} \left[\hH_s^{(n)} - \int d\vR \left(\vE(\vR, t)  +\frac{1}{\epsilon_0} \vP_{\parallel}^{(n)}(\vR) \right)\cdot \hP^{(n)}(\vR)\right]  
\end{aligned}
\end{equation}

\paragraph{Hamiltonians \# I'/ \# II'.}
Before presenting any results, one final point is appropriate. As discussed 
before, Eq. (\ref{eq:PZW_H}) should formally include the self-interaction of all charges. And, for a single electron at each site $n$, this self-interaction will be of the form $\hat{V}_{self} = \frac{1}{2\epsilon_0}\int d\vR |\hP_{\parallel}^{(n)}|^2$. If we make a semiclassical approximation (in the spirit of Eqs. (\ref{eq:Vdd_exp_1}) and (\ref{eq:Vdd})), we can approximate $\hat{V}_{self} = \frac{1}{\epsilon_0}\int d\vR \vP_{\parallel}^{(n)}\cdot \hP_{\parallel}^{(n)}$, which will obviously cancel the self-interaction terms in Eqs. (\ref{eq:Hsc_E_simulate}) and (\ref{eq:Hsc_D_simulate}). The resulting Hamiltonians will be of the form
\begin{subequations}
	\begin{align}
	\hH_{sc}^{I'} &=  \sum_{n=1}^{N} \left[\hH_s^{(n)} - \int d\vR \vE(\vR, t)  \cdot \hP^{(n)}(\vR)\right] 
	- \frac{1}{\epsilon_0}\sum_{n\neq l}\int d\vR\vP_{\parallel}^{(n)}(\vR) \cdot \hP^{(l)}(\vR) + \sum_{n <  l} \hV_{coul}^{(nl)}  \label{eq:Hsc_E_noself}\\
	\hH_{sc}^{II'} &=  \sum_{n=1}^{N} \hH_s^{(n)} - \int d\vR \vE(\vR, t)  \cdot \hP^{(n)}(\vR) \label{eq:Hsc_D_noself}
	\end{align}
\end{subequations}
In practice, as shown in the supporting information, we find that $\hH_{sc}^{I'}$  and $\hH_{sc}^{II'}$ behave effectively the same as  $\hH_{sc}^{I}$  and $\hH_{sc}^{II}$. In the supporting information, we list the relevant energy expression that is conserved for each choice of $\hH_{sc}$.

\paragraph{A comparison of the different Hamiltonians.}
When comparing Hamiltonians \#I and \#II, it is very important to emphasize that, although we have derived $\hH_{sc}^{II}$ by invoking the approximation in Eq. (\ref{eq:Vdd}),  $\hH_{sc}^{II}$ can also be derived directly from the PZW Hamiltonian. $\hH_{sc}^{II}$ should not be considered any less valid than $\hH_{sc}^{I}$; see supporting information.

Next, let us comment on the issues of electronic correlation and quantum entanglement. As far as quantum entanglement is concerned, with semiclassical electrodynamics,  there cannot be any strict quantum entanglement between electrons and photons because the EM field is treated classically. Nevertheless, even with Ehrenfest dynamics, there is some feedback from the electronic degrees of freedom to the photon field, and there is certainly some correlation between the boson field and the electronic state at any given time.\cite{Parandekar2006} A great deal of research has now shown that Ehrenfest equations of motion can sometimes yield the proper dynamics for fermionic subsystems coupled to bosonic baths (especially provided that one works with the correct initial conditions).\cite{Cotton2013,Cotton2014}

Let us now move our attention to electron-electron correlation. One the one hand, because Hamiltonian \#I contains a quantum two-body operator (i.e., $\hat{V}_{Coul}^{(nl)}$ in Eqs. (\ref{eq:Vdd_exp_1}-\ref{eq:Vdd_interaction_point})), this method allows for entanglement between individual molecules.
On the other hand, by invoking a classical intermolecular Coulomb operator in Eq. (\ref{eq:Vdd}), Hamiltonian \#II does not allow for entanglement between molecules.  As a practical matter, in what follows below, we will see that these differences can lead to different energy transfer rates.

To compare the two semiclassical Hamiltonians above, we will now apply Ehrenfest electrodynamics and model resonant energy transfer (RET) between  a pair of identical electronic two-level systems (TLSs)\cite{West1985, Spano1989, AlfonsoHernandez2015, Mirkovic2017} in three dimensions.

\paragraph{Model.}
Consider a pair of TLSs with a donor ($D$) and an acceptor ($A$). The Hamiltonian for both the donor and acceptor are 
\begin{equation}\label{eq:Hs}
\hH_s^{(D)} = \hH_s^{(A)} = 
\begin{pmatrix}
0 & 0 \\0 & \hbar\omega_0
\end{pmatrix}
\end{equation}
where Eq. (\ref{eq:Hs}) is expressed in the basis $\{\ket{g}, \ket{e}\}$; here $\ket{g}$ is the ground state and $\ket{e}$ is the excited state. $\hbar\omega_0$ is the energy difference between $\ket{g}$ and $\ket{e}$. The polarization operator for each molecule reads
\begin{equation}\label{eq:P}
\hP^{(n)}(\vR) = \vxi(\vR - \vR_0^{(n)})
\begin{pmatrix}
0 & 1 \\1 & 0
\end{pmatrix}, \ \ \ n = D, A
\end{equation}
Here, $\vxi(\vR) = \psi^{\ast}_{g}q\vR\psi_e =  (2\pi)^{-3/2}\sigma^{-5}\mu_{12}\vR z \exp(-r^2/2\sigma^2)$ is the polarization density of a TLS where $\ket{g}$ is an $s$-orbital, $\ket{e}$   is a $p_z$ orbital, $q$ denotes the effective charge of the TLS, $\sigma$ denotes the width of wave functions and $\mu_{12} = |\int d\vR \psi^{\ast}_{g}q\vR\psi_e|$ denotes the magnitude of transition dipole moment. We assume the TLS has no permanent dipole. Without loss of generality, we suppose  the donor (acceptor) sits on the negative (positive) side of the $x$-axis, i.e., $\vR_0^{(D)} = (-R/2, 0, 0)$ and $\vR_0^{(A)} = (R/2, 0, 0)$. We define  $R$ as the separation between the two TLSs. 

Overall, the electronic Hamiltonians read as follows in matrix form (in the basis \{$\ket{gg}$, $\ket{ge}$, $\ket{eg}$, $\ket{ee}$\}): 
\begin{equation}\label{eq:HI_matrix}
\hH_{sc}^{I} = 
\begin{pmatrix}
0 &v_A &v_D &v \\
v_A & \hbar\omega_0 & v & v_D\\
v_D & v & \hbar\omega_0 & v_A\\
v & v_D & v_A &2\hbar\omega_0
\end{pmatrix}
\end{equation}
and
\begin{equation}\label{eq:HII_matrix}
\hH_{sc}^{II} = 
\begin{pmatrix}
0 &v_A' &v_D' & 0 \\
v_A' & \hbar\omega_0 & 0 & v_D'\\
v_D' & 0 & \hbar\omega_0 & v_A'\\
0 & v_D' & v_A' &2\hbar\omega_0
\end{pmatrix}
\end{equation}
where $v = \frac{1}{\epsilon_0}\int d\vR \ \vxi^{(D)}_{\parallel}\cdot\vxi^{(A)}_{\parallel}$, $v_{D} = -\int d\vR \ \vE_{\perp}\cdot\vxi^{(D)}$ and $v_{D}' = -\frac{1}{\epsilon_0}\int d\vR \ \vxi^{(D)}\cdot\left(\vD - 2\text{Re}\rho_{12}^{(D)}\vxi^{(D)}_{\perp}\right)$, and $v_{A}$ and $v_{A}'$ are defined analogously.
All other simulation details and parameters are provided in the supporting information.

\paragraph{Analytical QED results.}
When modeling RET with retardation\cite{Andrews1992, Andrews1989, Andrews1989Unified}, it is well known that energy transfer rates show an $R^{-6}$ dependence when $k_0R \ll 1$ and an $R^{-2}$ dependence when $k_0R \gg 1$. Here $k_0 \equiv \omega_0 /c$.  This difference in scaling arises because the usual
instantaneous version of energy transfer theory\cite{Scholes2003, lakowicz2004principles, Nelson2013} does not account for the dynamical motion of the EM field to carry energy from donor to acceptor. 
For our purposes, in order to directly compare with simulation, we will require an accurate calculation of energy transfer dynamics (beyond any rate expression, e.g., F\"orster theory) that is exact within QED perturbation theory. A short-time analytical formula of the excited state population of the acceptor, $\rho_{22}^{(A)}(t)$, can be derived with QED, as shown by Power, Thirunamachandran and Salam\cite{Power1983-3,salam2010molecular}. By slightly modifying the result in Ref. \citenum{salam2010molecular},
we can obtain an analytical solution for $\rho_{22}^{(A)}(t)$  at short times,  starting in an arbitrary superposition state for the donor (see supporting information),
\begin{equation}\label{eq:P_2_Power}
\begin{aligned}
\rho_{22}^{(A)}(t) = & \frac{\rho_{22}^{(D)}(0)}{(4\pi\epsilon_0\hbar)^2}\left| \mu_{12}^{D}\mu_{12}^{A}\left[-\eta_1\frac{k_0^2}{R} + \eta_3\left(\frac{1}{R^3} - \frac{ik_0}{R^2}\right)\right] \right|^2  \\
& \times \left(t - \frac{R}{c}\right)^2 \theta\left(t-  \frac{R}{c}\right)
\end{aligned}
\end{equation}
Here, $\rho_{22}^{(D)}(0)$ is the initial excited state population of the donor, $\ve_D$ and $\ve_A$  are the unit vectors oriented along the transition dipoles of the donor and the acceptor, $\eta_{1} = \ve_A\cdot \ve_D - (\ve_A\cdot\ve_R)(\ve_D\cdot\ve_R)$ and $\eta_{3} = \ve_A\cdot \ve_D - 3(\ve_A\cdot\ve_R)(\ve_D\cdot\ve_R)$. We define $\ve_R$ as the unit vector  oriented along the separation between donor and acceptor. In our model,  the pair of  TLSs are located along the $x$-axis and the transition dipole moments are both $p_z$ polarized, so that $\ve_A\cdot\ve_R = \ve_D\cdot\ve_R=0$ and $\eta_1 = \eta_3 = \ve_A\cdot\ve_D = 1$. $\theta(t) = \frac{d}{dt}\text{Max}\{t, 0\}$ is the Heaviside step function.

Note that the unretarded energy transfer expression for $\rho_{22}^{(A)}$ is simply $\rho_{22}^{(A)}(t) =  \rho_{22}^{(D)}(0)\times \frac{|\mu_{12}^{D}|^2|\mu_{12}^{A}|^2}{(4\pi\epsilon_0\hbar)^2R^6} \eta_{3}^2t^2$, which is equivalent to the FGR result with the coupling $\hV_{Coul}^{(nl)}$ in Eq. (\ref{eq:Vdd_interaction_point}).
 Eq. (\ref{eq:P_2_Power}) includes two important time-dependent features: (i) all retardation is totally accounted for (i.e., $\rho_{22}^{(A)}(t)$ is zero when $t < R/c$) and (ii) $\rho_{22}^{(A)}(t)$ depends quadratically on time at short times.

\paragraph{Numerical semiclassical results.}
As far as simulating energy transfer semiclassically, 
we will assume that there is no EM field in space initially,
the donor starts in a superposition state $(C_1^{(D)}(0), C_2^{(D)}(0)) = (1/\sqrt{2}, 1/\sqrt{2})$  and the acceptor starts in the ground state, where $C_1$ ($C_2$) represents the quantum amplitude of $\ket{g}$ ($\ket{e}$).  With these initial conditions, we can propagate Eqs. (\ref{eq:Ehrenfest}), 
and compare dynamics of Hamiltonians \#I  and \#II. To keep the following context concisely, we will refer to the result of Hamiltonian \#I (II) as result \#I (II) for short.

\begin{figure*}
	\centering
	\includegraphics[width=0.8\textwidth]{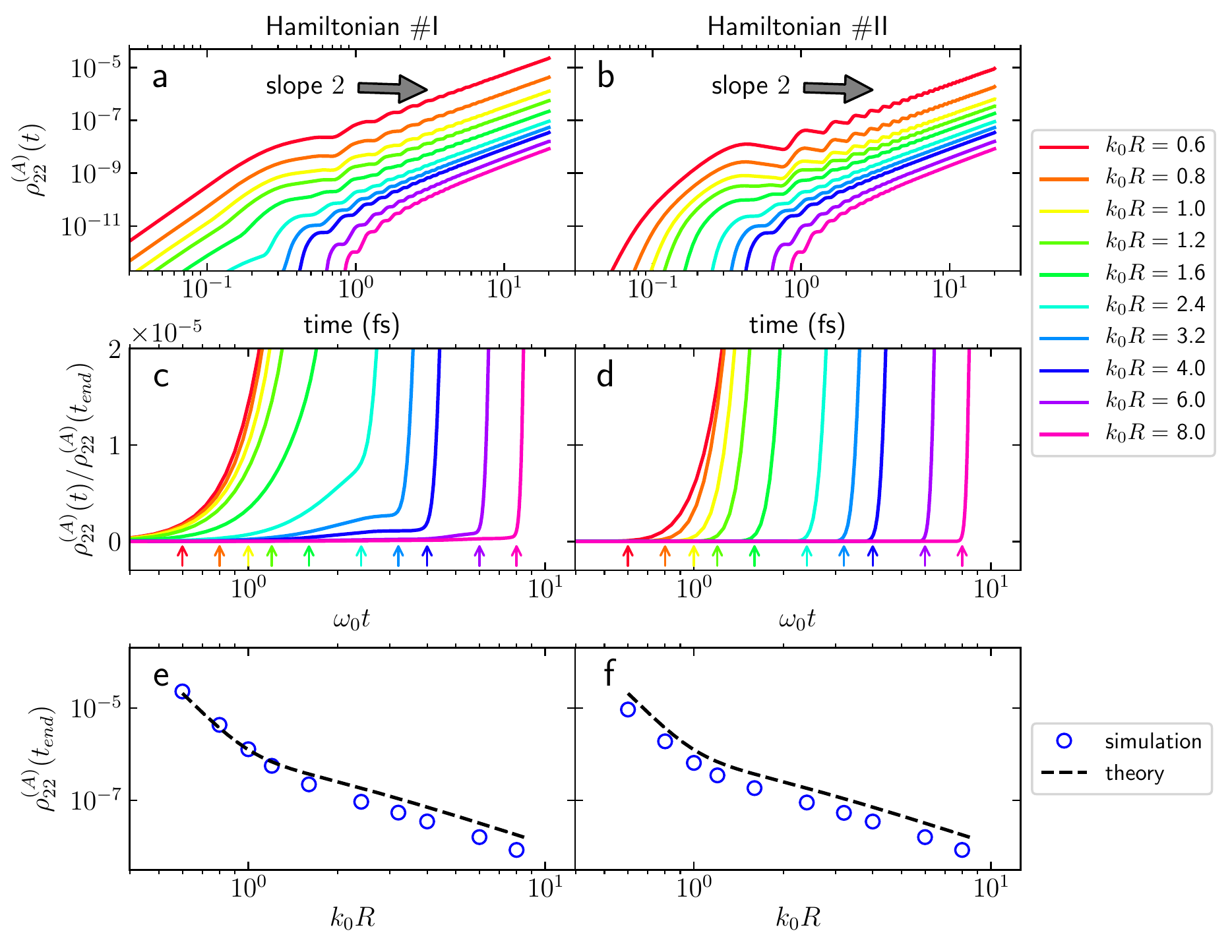}
	\caption{
		Plot of the excited state population of the acceptor ($\rho_{22}^{(A)}(t)$) at short times ($t_{end} = 20$ fs). Results for  Hamiltonian \#I (II) are plotted on the left (right). (a-b) $\rho_{22}^{(A)}(t)$ versus time using a  logarithmic scale by varying the separation in the range $0.6 \leq k_0R \leq 8.0$ (rainbow color from red to purple respectively), where $k_0 = \omega_0 / c$; (c-d) Normalized $\rho_{22}^{(A)}$ ($\rho_{22}^{(A)}(t) / \rho_{22}^{(A)}(t_{end})$) versus $\omega_0 t$ with the same separation range as in Fig. a-b, where now only the $x$-axis is plotted logarithmically; (e-f)  $\rho_{22}^{(A)}(t_{end})$ versus $k_0 R$ in logarithmic scale; the simulation data (blue circles) of Hamiltonians \#I and \#II are compared with the QED result (Eq. (\ref{eq:P_2_Power}), black dashed line) respectively. The initial state for the donor is $(C_1^{(D)}(0), C_2^{(D)}(0)) = (1/\sqrt{2}, 1/\sqrt{2})$ and the initial state for the acceptor is $(C_1^{(A)}(0), C_2^{(A)}(0)) = (1, 0)$. Other parameters are given in the supporting information. 
		Note that in Figs. a-b the straight lines when $t > 2$ fs  indicate that the leading term of $\rho_{22}^{(A)}(t)$ varies $\sim t^2$ (same as  Eq. (\ref{eq:P_2_Power})). Note that Hamiltonian \#I (Fig. c) violates causality such that  $\rho_{22}^{(A)}(t) > 0$ before the retarded field from the donor comes ($\omega_0 t < k_0R$) while Hamiltonian \#II (Fig. d) exactly preserves causality; see the rainbow arrows indicating the time before which energy transfer is not allowed by causality. In Figs. e-f, Both Hamiltonians show  $R^{-6}$ dependence when $k_0R < 1$ and  $R^{-2}$ dependence when $k_0 R > 1$. However, Hamiltonian \#I agrees with QED better for short separations than Hamiltonian \#II, presumably because the former describes Coulomb interactions quantum-mechanically.
	}
	\label{fig:1}
\end{figure*}

In Figs. \ref{fig:1}, we plot the excited state population of the acceptor  ($\rho_{22}^{(A)}(t)$) at relatively short times ($t < 20$ fs) by varying the separation $R$, ($0.6 \leq k_0 R\leq 8.0$). 
In Fig. \ref{fig:1}\textit{c}, we find that result  \#I  clearly doesn't preserve causality: $\rho_{22}^{(A)}(t)$ begins to increase
even before the retarded field from the donor arrives ($\omega_0 t < k_0 R$); see supporting information for a discussion of causality. 
Interestingly, however,  for very large distances (when $k_0 R \gg 1$),  Hamiltonian \#I
seems to do a better job of preserving causality because, in this limit, the  intermolecular interactions are dominated by the retarded field (which decays as $R^{-1}$) rather than 
longitudinal Coulomb interactions (which decay as $R^{-3}$).  Nevertheless, clearly, Hamiltonian \#I violates the tenets of relativity.  That being said,  
Hamiltonian \#II does preserve causality exactly (see Fig. \ref{fig:1}\textit{d}). Thus, from this perspective, one would presume
Hamiltonian \#II has an obvious advantage over Hamiltonian \#I.

At this point, however, let us turn our attention to Figs. \ref{fig:1}\textit{e-f}.  Here,  we compare rates of energy transfer for the two methods as compared with the analytic theory in Eq. (\ref{eq:P_2_Power}) as a function of $R$.   According to Fig. \ref{fig:1}\textit{e-f}, even though results  \#I and \#II (blue circles) recover qualitatively the same distance dependencies as   Eq. (\ref{eq:P_2_Power}) (black lines), results  \#I and \#II differ in the limit of short donor-acceptor separation ($k_0R < 1$). For short distances,
result \#I  agrees exactly with QED (Eq. (\ref{eq:P_2_Power})) while results  \#II  is off by roughly a factor of two. This discrepancy 
is perhaps not surprising because, at short separation, the dominant Coulomb interactions are described quantum-mechanically in Hamiltonian \#I but are classical in Hamiltonian \#II,
and there is no reason to suppose that these two methods should agree quantitatively in practice. By contrast, at long separations ($k_0R > 1$) -- where the retarded field is dominant -- both Hamiltonians \#I and \#II propagate the retarded field classically, and so both methods should agree; interestingly, in this limit, both semiclassical approaches 
differ  from the QED results by roughly a factor of two.\cite{footnoteNew}

\paragraph{Can we model energy transfer accurately without spontaneous emission?}
At large separation ($k_0R \gg 1$), it is clear that RET is dominated by the dynamics of the radiation field: retardation effects appear and the RET rate scales as $1/R^2$ instead of the usual $1/R^6$ scaling (i.e., the F\"orster scaling that arises from the instantaneous dipole-dipole interactions).  Now, for this reason, if semiclassical theory is to model RET correctly, it is clear that one must treat spontaneous emission correctly. After all, at long distances, RET can effectively be considered as the result of spontaneous emission from the donor, followed subsequently by absorption of the acceptor. That being said, however, we must emphasize  that
Ehrenfest electrodynamics do not recover the full FGR spontaneous emission rate\cite{schwabl2005, Miller1978, Li2018Spontaneous}. Instead, as shown in Ref.  \citenum{Li2018Spontaneous}, 
Ehrenfest dynamics predict a decay rate ($k_{Eh}$) proportional to the instantaneous ground state population: 
\begin{equation}\label{eq:kEh}
\kEh(t) = \rho_{11}(t)\kFGR
\end{equation}
One can argue that this failure arises from the fact that Ehrenfest electrodynamics predict only a coherent scattering field (which is proportional to the ground state population of the molecule) without any incoherent scattering \cite{Mollow1969, Chen2018Spontaneous}. In other words, according to a single Ehrenfest trajectory, one would predict $\avg{\hE}^2 = \avg{\hE^2}$, which is not correct quantum-mechanically.  By contrast, according to a quantum treatment, both coherent and incoherent scattering are allowed, and interference effects can lead to situations where, in the extreme case, $\avg{\hE} = 0$ but $\avg{\hE}^2 \neq 0$, as is  common for spontaneous emission. Thus, to sum up, modeling RET robustly requires more than a single classical ansatz for the electric field at one time, $\avg{\hE(t)}$:  a FGR calculation relies on capturing the correct time correlation function for the electric field, $\avg{\hE(0)\hE(t)}$; see note about averaging Ehrenfest trajectories in the supporting information.

\begin{figure*}
	\centering
	\includegraphics[width=\textwidth]{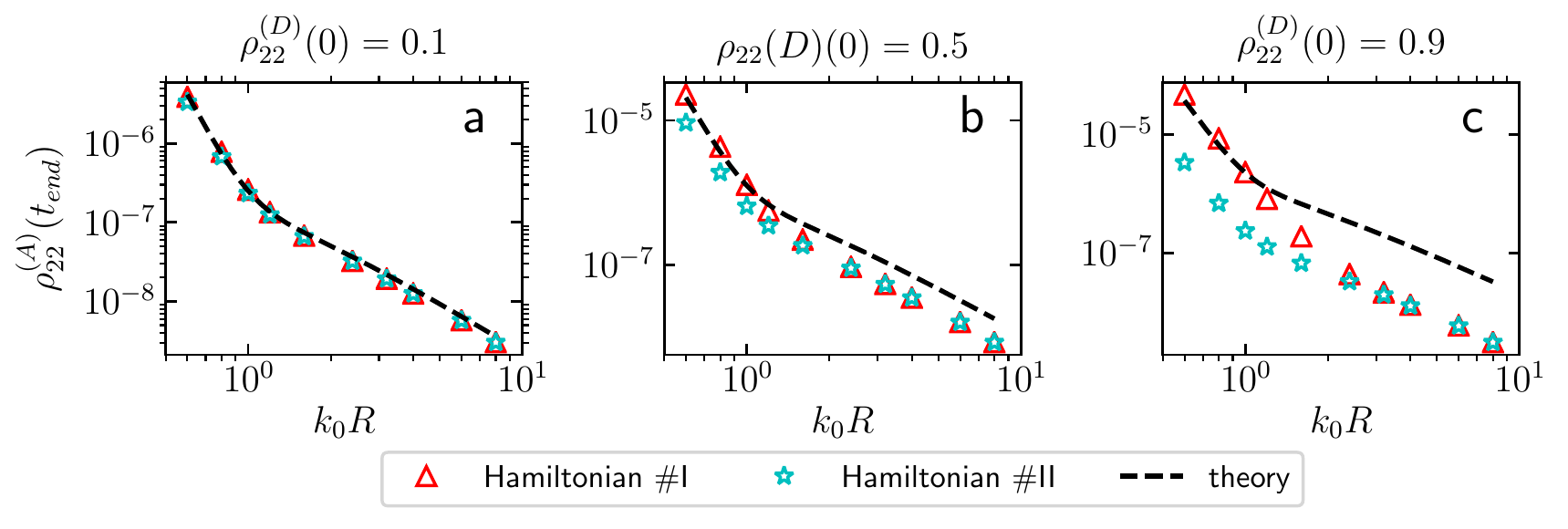}
	\caption{
	Plot of the excited state population of the acceptor at the end time ($\rho_{22}^{(A)}(t_{end})$, $t_{end} = 20$ fs) versus the intermolecular separation ($k_0 R$) using a  logarithmic scale. Simulations are performed with different initial excited state populations for the donor: $\rho_{22}^{(D)}(0)$ =  0.1 (left), 0.5 (middle) and 0.9 (right).  Three methods are compared: Hamiltonian \#I (red triangle), Hamiltonian \#II (cyan star) and QED (Eq. (\ref{eq:P_2_Power}), black dashed line). Parameters are given in the supporting information. Note that when $\rho_{22}^{(D)}(0)$ is small, all methods agree with each other. As $\rho_{22}^{(D)}(0)$ increases, there is less agreement between Hamiltonians \#I/II and the QED result. Just as for Fig. \ref{fig:1}, due to its quantum-mechanical description of Coulomb interactions,  Hamiltonian \#I always agrees with QED better for short separations (unlike Hamiltonian \#II).
	}
	\label{fig:2}
\end{figure*}

With this background and Eq. (\ref{eq:kEh}) in mind, one might expect that the Ehrenfest energy transfer rate would depend strongly on initial state population, and one can ask:
will our results using Hamiltonians \#I and \#II change in a similar fashion for different initial states?
To that end, in Figs. \ref{fig:2}, for a variety of initial conditions, we compare results for $\rho_{22}^{(A)}(t)$ as calculated according to both
Hamiltonians \#I (red triangle) and \#II (cyan star). We also plot the short time fully QED results (black dashed line) from Eq. (\ref{eq:P_2_Power}),
where the initial excited state population is reflected in the initial donor ($\rho_{22}^{(D)}(0)$). 

Our results are plotted in Figs. \ref{fig:2}.  When the donor is weakly excited initially ($\rho_{22}^{(D)}(0) = 0.1$), we find that all three results agree with each other.  However, when $\rho_{22}^{(D)}(0)$ is increased, we find less and less agreement between either of the semiclassical results and QED results {\em at long distances};
the semiclassical results strongly underestimate the energy transfer rate.   These results strongly suggest that, if a semiclassical approach is to capture energy transfer accurately both at short and long distances, the approach must be able to capture  
spontaneous emission as well.  After all, at long distances, we know that energy transfer is modulated by a retarded field, and if Ehrenfest dynamics cannot capture spontaneous emission, there is no surprise that one cannot recover the correct energy transfer rate either.  

Lastly, let us now consider results at short distances. Here, we find very different behavior between Hamiltonians \#I and \#II. 
On the one hand, we find that,  no matter the initial donor population, Hamiltonian \#I always produces accurate results; 
because Hamiltonian  \#I includes explicitly quantum-mechanical Coulomb interactions, we believe this method should always agree
with QED at short range (where retardation effects are not important).
 On the other hand, in Fig. \ref{fig:2}\textit{c}, we also see that 
Hamiltonian \#II fails and drastically underestimates the energy transfer rate for $\rho_{22}^{(D)}(0) = 0.9$. 
Here, we need only recognize that, 
because Hamiltonian \#II treats the EM field exclusively classically, such an approach can never be accurate (either at short range or at long range)
if spontaneous emission is not capture correctly. Thus, in the end, a crucial question emerges: If we can develop a means to include
spontaneous emission on top of Ehrenfest dynamics (as in Ref. \citenum{Chen2018Spontaneous}), what will be the most accurate approach: to include a combination of quantum Coulomb interactions with
a classical (but exclusively transverse) EM field (i.e. Hamiltonian \#I)? Or to employ an entirely classical (transverse plus longitudinal) EM field?  
The answer is not obvious, especially because the full nature of a quantum radiation field cannot be captured by simply including spontaneous emission. Hence, a thorough benchmark will be necessary.
As we look forward to future methodological development of this understudied area, many questions remain.


In conclusion, by numerically studying coherent energy transfer between a pair of TLSs with Ehrenfest electrodynamics, our conclusions are as follows. $(i)$ The standard Hamiltonian \#I ($\hH_{sc}^{I}$ in Eq. (\ref{eq:Hsc-Escheme})) violates causality, especially when the molecular separation is small ($k_0R < 1$) because of a mismatch between a \textit{quantum} description of the matter and a \textit{classical} description of the EM field; $(ii)$ Causality can be preserved if one models both the retarded field and the intermolecular Coulomb interactions in a classical fashion (Hamiltonian $\hH_{sc}^{II}$ in Eq. (\ref{eq:Hsc-Dscheme})).  $(iii)$ For RET, both Hamiltonians \#I and \#II predict qualitatively the same distance behavior as retarded F\"orster theory, and when the electronic excitation of the donor is weak, both semiclassical methods
recover QED results quantitatively.
However,  $(iv)$ even though Hamiltonian \#I violates causality, this approach better agrees with QED 
as far as RET rates at short distances.
The pros and cons of these different Hamiltonians suggest that the specific choice of a semiclassical Hamiltonian may depend on the particular problem one 
is investigating --- for now, it would appear
 there is no sinecure for the inconsistencies inevitably faced by a semiclassical ansatz.  Nevertheless, if spontaneous emission can be incorporated
into Ehrenfest dynamics, the accuracy of these methods should be dramatically enhanced. This work is ongoing in our laboratory.

\begin{acknowledgement}
 	This material is based upon work supported by the the National Science Foundation under Grant No. CHE-1764365 (T.L., H.T.C., J.E.S.), AFOSR Grant No. FA9550-15-1-0189 (M.S.), US-Israel Binational Science Foundation Grant No. 2014113 (M.S. and A.N.) and the US National Science Foundation Grant No. CHE1665291 (A.N.). 
\end{acknowledgement}

\begin{suppinfo}
	\begin{itemize}
		\item Derivation of semiclassical Hamiltonians; simulation details and parameters; the conserved quantity for each semiclassical Hamiltonian discussed above; results for Hamiltonians \#I' and \#II', showing that self-interaction is not important for accurately modeling the dynamics of energy transfer; details regarding the derivation of Eq. (\ref{eq:P_2_Power}) and a comment on applying Weisskopf-Wigner theory to RET; a brief discussion of causality in Fig. \ref{fig:1}; a more detailed  explanation for why we must average over many  Ehrenfest trajectories (rather than a single trajectory).
	\end{itemize}
\end{suppinfo}

\bibliographystyle{achemso}

\providecommand{\latin}[1]{#1}
\makeatletter
\providecommand{\doi}
{\begingroup\let\do\@makeother\dospecials
	\catcode`\{=1 \catcode`\}=2 \doi@aux}
\providecommand{\doi@aux}[1]{\endgroup\texttt{#1}}
\makeatother
\providecommand*\mcitethebibliography{\thebibliography}
\csname @ifundefined\endcsname{endmcitethebibliography}
{\let\endmcitethebibliography\endthebibliography}{}

\newpage

\begin{center}
	\textbf{\large Supplemental Information: A Necessary Trade-off for Semiclassical Electrodynamics:  Accurate Short-Range Coulomb Interactions versus the Enforcement of Causality?}
\end{center}

\setcounter{equation}{0}
\setcounter{figure}{0}
\setcounter{table}{0}
\setcounter{page}{1}
\makeatletter
\renewcommand{\theequation}{S\arabic{equation}}
\renewcommand{\thefigure}{S\arabic{figure}}
\renewcommand{\bibnumfmt}[1]{[S#1]}
\renewcommand{\citenumfont}[1]{S#1}

\section{Deriving Hamiltonians \#I and \#II}
To derive Hamiltonians \#I and \#II, we will largely follow Mukamel\cite{Mukamel1999}.
The PZW Hamiltonian in Eq. (\ref{eq:PZW_H}) can be separated into three parts:
\begin{equation}\label{H_PZW_total}
\hH_{PZW} = \hH_{M} + \hH_{R} +\hH_{I}
\end{equation} 
where $\hH_{M}$, $\hH_{R}$ and $\hH_{I}$ are Hamiltonians for the molecular subsystem, radiation field and the coupling between field and matter. They are defined as
\begin{subequations}\label{eq:H_PZW_separate}
	\begin{align}
	\hH_{M} &= \hH_s + \frac{1}{2\epsilon_0}\int d\vR \ |\hPperp(\vR)|^2 \label{eq:H_PZW_sep_HM}\\
	\hH_{R} &= \frac{1}{2}\int d\vR\ \left[ \frac{|\hDperp(\vR)|^2}{\epsilon_0} + \frac{|\hB(\vR)|^2}{\mu_0} \right]  \\
	\hH_{I} &= - \int d\vR\ \frac{\hDperp(\vR)}{\epsilon_0}\hPperp(\vR)
	\end{align}
\end{subequations}
$\hH_{s}$, $\hPperp$, $\hDperp$ and $\hB$ are defined in the main text. Note that $\hH_{s}$ and $\hPperp$ are operators acting on the molecular (matter) subspace while $\hDperp$ and $\hB$ are operators acting on the radiation field.
For an operator (say, $\hat{F}$) acting solely on the radiation field, by utilizing the Heisenberg equations of motion, one obtains $\frac{d}{dt}\hat{F} = \frac{i}{\hbar}\left[\hH_{R} + \hH_{I}, \hat{F}\right]$. Similarly, for an operator $\hat{Q}$  acting solely on the matter side, one obtains $\frac{d}{dt}\hat{Q} = \frac{i}{\hbar}\left[\hH_{M} + \hH_{I}, \hat{Q}\right]$. Applying Eq. (\ref{eq:H_PZW_separate}) and after some straightforward algebra, one finally obtains the equations of motion for both field and matter operators\cite{Power1983-1,Mukamel1999}:
\begin{subequations}\label{eq:EOM_quantum}
	\begin{align}
	\frac{\partial}{\partial t} \hat{Q} &= \frac{i}{\hbar}\left[\hH_{s}, \hat{Q}\right] - \frac{i}{2\hbar}\int d\vR \ \left(\hEperp(\vR, t)\left[\hP(\vR, t), \hat{Q}\right] + \left[\hP(\vR, t), \hat{Q}\right]\hEperp(\vR, t)\right)\label{eq:EOM_Q}\\
	\frac{\partial}{\partial t} \hB(\vR, t) &= - \nabla \times \hE(\vR, t) \label{eq:EOM_E}\\
	\frac{\partial}{\partial t} \hE(\vR, t) &= c^2 \nabla \times \hB(\vR, t) - \frac{\hat{\vJ}(\vR, t)}{\epsilon_0} \label{eq:EOM_B}
	\end{align}
\end{subequations}
Here, $\hE = \frac{1}{\epsilon_0}\left(\hD -\hP\right)$ is a joint operator for both the field and matter sides, and $\hat{\vJ} = \frac{\partial}{\partial t}\hP$. Eqs. (\ref{eq:EOM_quantum}) describe the full quantum dynamics for both the field and matter operators. Because these operators are in an infinite-dimensional Hilbert space, it's almost impossible to solve Eqs. (\ref{eq:EOM_quantum}) directly. To make progress, we will invoke the Ehrenfest approximation in the next part.

\subsection{Ehrenfest approximation}
Suppose the density operator for the whole system ($\hrho_t$) can be expressed as $\hrho_t = \hrho_M \otimes \hrho_R$, where $\hrho_M$ and $\hrho_R$ are the density operators of the molecular subsystem and radiation field. Note that this separation is allowable only when the electron and photon are not in an entangled state. 

In order to propagate EM fields in a real-space grid, the field operators should be reduced to classical variables, i.e.,  we use the total density operator $\hrho_t$ to trace over Eqs. (\ref{eq:EOM_E}-\ref{eq:EOM_B}), which leads to
\begin{subequations}\label{eq:EOM_EM_reduced}
	\begin{align}
	\frac{\partial}{\partial t} \avg{\hB(\vR, t)}_t &= -\avg{ \nabla \times \hE(\vR, t)}_t \label{eq:EOM_E_trace}\\
	\frac{\partial}{\partial t} \avg{\hE(\vR, t)}_t &= c^2 \avg{\nabla \times \hB(\vR, t)}_t - \frac{\avg{\hat{\vJ}(\vR, t)}_t}{\epsilon_0} \label{eq:EOM_B_trace}
	\end{align}
\end{subequations}
Here,  $\avg{\cdots}_t$ is short-hand for $\tr{\hrho_t\cdots}$. Since $\hat{\vJ}$ is an operator acting only on the molecular subsystem, $\avg{\hat{\vJ}(\vR, t)}_t = \frac{\partial}{\partial t}\tr{\hrho_M \hP(\vR, t)}$. Further assuming $\avg{\nabla\times\cdots}_t = \nabla\times \avg{\cdots}_t$ in the spirit of Ehrenfest dynamics, Eqs. (\ref{eq:EOM_EM_reduced}) are simplified to
\begin{subequations}\label{eq:EOM_EM_final}
	\begin{align}
	\frac{\partial}{\partial t} \vB(\vR, t) &= - \nabla \times \vE(\vR, t)\label{eq:EOM_E_final}\\
	\frac{\partial}{\partial t} \vE(\vR, t) &= c^2 \nabla \times \vB(\vR, t) - \frac{\vJ(\vR, t)}{\epsilon_0} \label{eq:EOM_B_final}
	\end{align}
\end{subequations}
where we define $\vE = \avg{\hE}_t$, $\vB = \avg{\hB}_t$ and $\vJ = \avg{\hat{\vJ}}_t = \frac{\partial}{\partial t}\tr{\hrho_M \hP(\vR, t)}$. Eqs. (\ref{eq:EOM_EM_final}) are the classical Maxwell's equations.

For the matter side, we seek a quantum-mechanical propagator, and so we trace Eq. (\ref{eq:EOM_Q}) by the reduced density operator for the EM field $\hrho_R$, which leads to
\begin{equation}\label{eq:EOM_Q_traced}
\frac{\partial}{\partial t} \hat{Q} = \frac{i}{\hbar}\left[\hH_{s}, \hat{Q}\right] - \frac{i}{2\hbar}\int d\vR \  \left(\avg{\hEperp(\vR, t)}_R\left[\hP(\vR, t), \hat{Q}\right] + \left[\hP(\vR, t), \hat{Q}\right]\avg{\hEperp(\vR, t)}_R\right)
\end{equation}
Note that since $\hE$ is a joint operator acting on both field and matter variables, $\avg{\hEperp}_R$ in Eq. (\ref{eq:EOM_Q_traced})  is still an operator for the matter, while $\avg{\hE}_t = \vE$ in Eqs. (\ref{eq:EOM_EM_reduced}-\ref{eq:EOM_EM_final}) is a classical variable. By further defining
\begin{equation}\label{eq:delta_E}
\hdelta = \avg{\hEperp}_R - \avg{\hEperp}_t = \avg{\hEperp}_R - \vE_{\perp}
\end{equation}
we can rewrite Eq. (\ref{eq:EOM_Q_traced}) as
\begin{equation}\label{eq:EOM_Q_final}
\begin{aligned}
\frac{\partial}{\partial t} \hat{Q} =& \ \frac{i}{\hbar}\left[\hH_s - \int d\vR \ \vE_{\perp}(\vR, t) \cdot \hP(\vR, t), \hat{Q}\right] \\
&- \frac{i}{2\hbar}\int d\vR \
\left(\hdelta\left[\hP(\vR, t), \hat{Q}\right] + \left[\hP(\vR, t), \hat{Q}\right]\hdelta\right)
\end{aligned}
\end{equation}
In Eq. (\ref{eq:EOM_Q_final}),  the equation of motion for the matter operator $\hat{Q}$ contains one Hamiltonian term (see the first line) and a second non-Hamiltonian, symmetrized term (see the second line). This second term is very interesting. On the one hand, this term reminiscent of the quantum-classical Liouville equation (QCLE)\cite{kapral1999mixed, kapral2006progress}. On the other hand, because $\avg{\hdelta}_M = 0$ and $\avg{\hdelta^2}_M \geq 0$, this term also resembles the noise in a Langevin equation for describing Brownian motion. Thus, the terms on the second line are crucial for recovering the correct quantum behavior behind both spontaneous emission and energy transfer. However, including the dynamics of $\hdelta$ are obviously difficult.

To make progress, the simplest strategy is to neglect the second term of Eq. (\ref{eq:EOM_Q_final}). By further using $\hrho_M$ (the reduced density operator for the matter side), we can reduce Eq. (\ref{eq:EOM_Q_final}) to
\begin{equation}\label{eq:EOM_Q_Ehrenfest}
\frac{\partial}{\partial t} \hrho_M = \ -\frac{i}{\hbar}\left[\hH_{s} - \int d\vR \ \vE_{\perp}(\vR, t) \cdot \hP(\vR, t), \hrho_M\right] 
\end{equation}
Eqs. (\ref{eq:EOM_EM_final}) and (\ref{eq:EOM_Q_Ehrenfest}) are the standard Ehrenfest electrodynamics derived by Mukamel.

Note that this simplification is only valid when $\avg{\hdelta^2}_M \ll \vE^2$. In the case of Brownian motion, the corresponding noise term is not negligible so that we cannot make of such a simplification. However, for light-matter interactions, if a molecule is weakly excited, the scattering field is dominated by coherent scattering, which means $\avg{\hdelta^2}_M \ll \vE^2$. Hence,  standard Ehrenfest electrodynamics should be valid. This conclusion can be numerically confirmed by Fig. \ref{fig:2} for the case of resonant energy transfer; Ref. \citenum{Li2018Spontaneous} discusses why, in the weak excitation limit, spontaneous emission emerges without including the second term.

\subsection{Deriving Hamiltonian \#I}
From the Ehrenfest electrodynamics defined in Eqs. (\ref{eq:EOM_EM_final}) and (\ref{eq:EOM_Q_Ehrenfest}), the semiclassical Hamiltonian obviously reads
\begin{equation}\label{eq:Hsc_Ehrenfest}
\hH_{sc} = \hH_{s} - \int d\vR \ \vE_{\perp}(\vR, t) \cdot \hP(\vR, t)
\end{equation}

For a system containing $N$ separable and neutral molecules, $\hH_s$ and $\hP$ can be expressed as in Eqs. (\ref{eq:many_particle_expression}), leading to Eq. (\ref{eq:Hsc-Escheme}) in the main body of the letter.

\subsection{Deriving Hamiltonian \#II}
To derive Hamiltonian \#II  directly from the PZW Hamiltonian, one can follow a very similar approach as above, only now one works with the PZW Hamiltonian in Eq. (\ref{eq:PZW_H_2}), for which one has already canceled out the instantaneous Coulomb interactions. Hence,  the Hamiltonian for the matter side reads
\begin{equation}
\hH_{M} = \sum_{n=1}^{N}\hH_s^{(n)} + \sum_{n=1}^{N}\frac{1}{2\epsilon_0}\int d\vR \ |\hPperp^{(n)}(\vR)|^2
\end{equation}
where $n$ is the index for molecules and $N$ is the total number of molecules. The definitions of $\hH_{R}$ and $\hH_{I}$ are the same as in Eqs. (\ref{eq:H_PZW_separate}). At this point, we follow the procedure from Eq. (\ref{eq:H_PZW_separate}) to (\ref{eq:EOM_Q_Ehrenfest}), and the semiclassical Hamiltonian in Eq. (\ref{eq:Hsc_II_Ehrenfest}) can be derived:
\begin{equation}\label{eq:Hsc_II_Ehrenfest}
\hH_{sc} =  \sum_{n=1}^{N}\hH_s^{(n)} + \sum_{n=1}^{N}\frac{1}{2\epsilon_0}\int d\vR \ |\hPperp^{(n)}(\vR)|^2 - \int d\vR \ \vD_{\perp}(\vR, t) \cdot \hP(\vR, t)
\end{equation}
Since we are limited to neutral molecular system, $\vD_{\perp} = \vD$. Eq. (\ref{eq:Hsc_II_Ehrenfest}) is very close to Hamiltonian \#II defined in Eq. (\ref{eq:Hsc-Dscheme}) expect for self-interaction term,  $\sum_{n=1}^{N}\frac{1}{2\epsilon_0}\int d\vR \ |\hPperp^{(n)}(\vR)|^2$ in Eq. (\ref{eq:Hsc_II_Ehrenfest}). This  self-interaction term should be compared with the term $\sum_{n=1}^{N}\frac{1}{\epsilon_0}\int d\vR \vP_{\perp}^{(n)}(\vR, t)\cdot \hP^{(n)}(\vR)$ in Eq. (\ref{eq:Hsc-Dscheme}). Overall, we believe these self-interaction terms will not  influence the dynamics of energy transfer and numerical evidence is  given later in the SI. Hence, we have now derived Hamiltonian \#II directly from the PZW Hamiltonian.

\subsection{Deriving Hamiltonians \#I and \#II from a Meyer-Miller-Stock-Thoss Mapping}
Other approaches are also possible for deriving Hamiltonian \#I (starting from the PZW Hamiltonian defined in Eq. (\ref{eq:PZW_H})), or Hamiltonian \#II (starting from the alternative expression for the PZW Hamiltonian in Eq. (\ref{eq:PZW_H_2})). For example, one of the most powerful methods for deriving Ehrenfest dynamics is to invoke a  Meyer-Miller-Stock-Thoss (MMST) mapping\cite{meyera1979classical,stock1997semiclassical}, which connects a nonadiabatic problem with $F$ bosonic degrees of freedom (DOFs) and $n$ fermonic DOFs into an adiabatic problem with $F+n$ bosonic DOFs. According to MMST, one expresses the electronic DOFs in the form of action-angle variable, converts these variables into harmonic oscillator creation and annihilation operators ($\hat{a}$, $\hat{a}^{\dagger}$), and then substitutes in the expressions $\hat{a}=\hat{x}+i\hat{p}$ and $\hat{a}^{\dagger}=\hat{x}-i\hat{p}$. Up to this point, the transformation is exact. If one now further treats $\hat{x}$ and $\hat{p}$ classically and on the same footing as $\vE$ and $\vB$, one can derive Ehrenfest dynamics. For the present case, one can recover both Hamiltonians \#I and \#II.

As a side note, we mention that classical MMST dynamics are equivalent to Poisson bracket mapping equation (PBME) dynamics, as constructed by Kapral and coworkers as an approximation to the  QCLE.\cite{kim2008quantum,nassimi2010analysis} Future work understanding Ehrenfest dynamics in the  context of the QCLE may also be very fruitful.

\section{Simulation details and parameters}\label{sec:Appendix}
We evolve the EM field and the total density matrix $\rho = \rho^{(D)} \otimes \rho^{(A)}$ by Eqs. (\ref{eq:Ehrenfest}). Maxwell's equations are simulated using the finite-difference time-domain (FDTD) method\cite{Taflove1998} with the perfect matching layer (PML) as the absorbing boundary condition. We use an open-source package \textit{MIT Electromagnetic Equation Propagation\cite{Oskooi2010}} (Meep) to propagate the EM field and update the current source $\vJ$ at each time step through Eq. (\ref{eq:J}). We propagate $\rho$ by $\rho(t+\Delta t) = \exp(-i\hH_{sc}(t) \Delta t /\hbar)\rho(t)$, where $\hH_{sc}$ is either calculated as Hamiltonian \#I (Eq. (\ref{eq:Hsc_E_simulate})) or \#II (Eq. (\ref{eq:Hsc_D_simulate})). Note that in practice, we do not directly calculate the time-consuming transverse E-field in Eq. (\ref{eq:Hsc-Escheme}) at each time step. Instead, 
because the polarization operators can be separated into a product of the spatial and  electronic parts, we calculate the spatial integrals $\frac{1}{\epsilon_0}\int d\vR\vP_{\parallel}^{(n)}(\vR) \cdot \hP^{(l)}(\vR)$ and  $\frac{1}{\epsilon_0}\int d\vR\vP_{\parallel}^{(n)}(\vR) \cdot \hP^{(n)}(\vR)$ 
{\em only one time}  (for each $R$) at the beginning of the calculation. To calculate the longitudinal vector field, we transform to Fourier space, where $\vP_{\parallel}(\vk) = \vk\left[\vk \cdot \vP(\vk)\right]/ |\vk|^2$, and then take the inverse Fourier transform.
For calculating $\hH_{sc}^{II}$, we use Eq. (\ref{eq:Hsc_D_simulate}) above.
As shown in the supplementary information, $\hH_{sc}^{I'}$ ($\hH_{sc}^{II'}$) behaves exactly the same as $\hH_{sc}^{I}$ ($\hH_{sc}^{II}$), suggesting that the term $\frac{1}{\epsilon_0}\int d\vR \vP_{\parallel}^{(n)}\cdot\hP_{\parallel}^{(n)}$ is not very important for energy transfer dynamics.

The parameters for the TLSs are as follows: for the transition dipole moment $\mu_{12} = 9.57\times 10^4 \text{ C}\cdot\text{nm}/\text{mol}$, for  the energy difference $\hbar\omega_0=6.58\text{ eV}$, and for the molecular width $\sigma = 3 \text{ nm}$. In the FDTD simulation, we calculate the EM field in a $(96+R) \text{\ nm} \times 96 \text{\ nm} \times 96 \text{\ nm}$ grid with spacing $\Delta x = 3 \text{ \  nm}$, where $R$ is the separation between two TLSs. We choose a small time step $\Delta t = 2 \times 10^{-4} \text{\ fs}$ to guarantee the accuracy and convergence. 

Note that for our simulations, we set  $\vE$ and $\vB$ initially to be exactly zero throughout space. Of course a more accurate initial condition would be to set $\vD$ 
equal to $\mathbf{0}$ everywhere (instead of $\vE$), given that the QED calculation are performed assuming no excitation of the D-photon field. Nevertheless, for RET, since $\vD(0) = \epsilon_0\vE(0) + \vP^{(D)}(0)$ and $\vP^{(D)}(0)\sim\text{Re}\rho_{12}^{(D)}(0)$, setting $\vE$ equal to $\mathbf{0}$ should not change the overall population dynamics much if we average over initial phases of $\rho_{12}$. Moreover, the initial energy in the field ($\frac{1}{2\epsilon_0} \int d\vR\left| \vP^{D}(0) \right|^2$) is very small compared to $\hbar\omega_0$, so these initial conditions should not be very different in practice.

\section{Energy conservation}\label{sec:Appendix2}
For each semiclassical Hamiltonian discussed in the manuscript, we can define a total energy function that is conserved. These energies are as follows:

\begin{table}
	\caption{Energy conservation for semiclassical Hamiltonians}
	\label{tbl:energy_conservation}
	\scriptsize
	\centering
	\begin{tabular}{lcr}
		\hline
		Hamiltonian & $\hH_s^{total}$ & conserved quantity\\
		\hline
		\# I (Eqs. (\ref{eq:Hsc-Escheme}-\ref{eq:Hsc_E_simulate}))& $\sum\limits_{n=1}^{N}\hH_s^{(n)} +  \sum\limits_{n <  l}\hV_{Coul}^{(nl)}$ &$\frac{1}{2}\int d\vR \left(\epsilon_0|\vE_{\perp}|^2+  
		\frac{|\vB|^2}{\mu_0} \right)+ \tr{\rho\hH_s^{total}}$ \\
		& & = $\frac{1}{2}\int d\vR \left(\epsilon_0|\vE_{\perp}|^2+  
		\frac{|\vB|^2}{\mu_0} \right)+ \tr{\rho\hH_s} +\frac{1}{\epsilon_0} \sum\limits_{n <  l}\vP_{\parallel}^{(n)}\cdot\vP_{\parallel}^{(l)}$
		\\ \hline
		\# II  (Eqs. (\ref{eq:Hsc-Dscheme}-\ref{eq:Hsc_D_simulate}) )& $\sum\limits_{n=1}^{N}\hH_s^{(n)} - \frac{1}{\epsilon_0}\int d\vR \vP_{\parallel}^{(n)}\cdot\hP_{\parallel}^{(n)}$ &$\frac{1}{2}\int d\vR \left(\epsilon_0|\vE|^2+  \frac{|\vB|^2}{\mu_0} \right) + \tr{\rho\hH_s^{total}} + \frac{1}{2\epsilon_0}\sum\limits_{n=1}^{N}\int d\vR |\vP_{\parallel}^{(n)}|^2$\\ 
		& &  
		= $\frac{1}{2}\int d\vR \left(\epsilon_0|\vE_{\perp}|^2+  \frac{|\vB|^2}{\mu_0} \right) + \tr{\rho\hH_s} +\frac{1}{\epsilon_0} \sum\limits_{n <  l}\vP_{\parallel}^{(n)}\cdot\vP_{\parallel}^{(l)}$
		\\ \hline
		\# I'  (Eq. (\ref{eq:Hsc_E_noself}))& $\sum\limits_{n=1}^{N}\hH_s^{(n)}+  \sum\limits_{n <  l}\hV_{Coul}^{(nl)} - \frac{1}{\epsilon_0}\int d\vR \vP_{\parallel}^{(n)}\cdot\hP_{\parallel}^{(n)}$ & $\frac{1}{2}\int d\vR \left(\epsilon_0|\vE|^2+  \frac{|\vB|^2}{\mu_0} \right) + \tr{\rho\hH_s^{total}} + \frac{1}{2\epsilon_0}\sum\limits_{n=1}^{N}\int d\vR |\vP_{\parallel}^{(n)}|^2 $\\ 
		& & 
		$ = \frac{1}{2}\int d\vR \left(\epsilon_0|\vE_{\perp}|^2+  \frac{|\vB|^2}{\mu_0} \right) + \tr{\rho\hH_s}  +\frac{1}{\epsilon_0} \sum\limits_{n \neq  l}\vP_{\parallel}^{(n)}\cdot\vP_{\parallel}^{(l)}$
		\\ \hline
		\# II' (Eq. (\ref{eq:Hsc_D_noself}))& $\sum\limits_{n=1}^{N}\hH_s^{(n)} $ &  $\frac{1}{2}\int d\vR \left(\epsilon_0|\vE|^2+  
		\frac{|\vB|^2}{\mu_0} \right)+ \tr{\rho\hH_s^{total}}$ \\
		& & 
		= $\frac{1}{2}\int d\vR \left(\epsilon_0|\vE_{\perp}|^2+  
		\frac{|\vB|^2}{\mu_0} \right)+ \tr{\rho\hH_s} + \frac{1}{2\epsilon_0}\int d\vR |\vP_{\parallel}|^2$
		\\ \hline
	\end{tabular}
	\normalsize
\end{table}
In Table \ref{tbl:energy_conservation}, $\hH_s = \sum\limits_{n=1}^{N}\hH_s^{(n)} $.

\section{Self-interaction is not important in energy transfer}\label{sec:Sself-interaction}
Here, we study the different approximations for the electronic Hamiltonian as discussed in the manuscript. First, we will show that
the  results for Hamiltonians \#I and \#II are barely changed if we ignore self-interaction, i.e., we ignore all terms of the form $\frac{1}{\epsilon_0}\int d\vR\ \vP_{\parallel}^{(n)}\cdot\hP^{(n)}$ in Eqs. (\ref{eq:Hsc_E_simulate}) and (\ref{eq:Hsc_D_simulate}). With this approximation, Hamiltonian \#I in Eq. (\ref{eq:Hsc_E_simulate}) becomes $\hH_{sc}^{I'}$ in Eq. (\ref{eq:Hsc_E_noself}), and 
Hamiltonian \#II in Eq. (\ref{eq:Hsc_D_simulate}) becomes $\hH_{sc}^{II'}$ in Eq. (\ref{eq:Hsc_D_noself}).

Alternatively, since Hamiltonian \#II can also be expressed as Eq. (\ref{eq:Hsc-Dscheme}), if we ignore the self-interaction term $ \frac{1}{\epsilon_0}\int d\vR \vP_{\perp}^{(n)}(\vR, t)\cdot \hP^{(n)}(\vR)$ in Eq. (\ref{eq:Hsc-Dscheme}), we obtain yet another slightly different Hamiltonian,  $\hH_{sc}^{II''} $
\begin{equation}\label{eq:Hsc-Dscheme_noself}
\begin{aligned}
\hH_{sc}^{II''} = \sum_{n=1}^{N} \hH_s^{(n)} - \frac{1}{\epsilon_0}\int d\vR \vD(\vR, t)\cdot \hP^{(n)}(\vR) 
\end{aligned}
\end{equation}  
For Hamiltonian \# II'',  the most natural conserved energy function is simply $\frac{1}{2}\int d\vR\left (\epsilon_0|\vE|^2+   
\frac{|\vB|^2}{\mu_0} \right)+ \tr{\rho\hH_s^{total}} - \frac{1}{2\epsilon_0}\int d\vR |\vP|^2$, where $\hH_s^{total} = \sum_{n=1}^{N}\hH_s^{(n)}$.

Second, we will study whether or not one can model resonant energy transfer accurately if the intermolecular interactions are described by the total E-field $\vE$ together with the  intermolecular Coulomb interactions $\hV_{coul}^{(nl)}$. This Hamiltonian obviously double counts but what are the practical effects of such double counting? This leads us to model Hamiltonian \#III,
\begin{equation}\label{eq:Hsc_EV}
\begin{aligned}
\hH_{sc}^{III} =  \sum_{n=1}^{N} \hH_s^{(n)} - \int d\vR \vE(\vR, t)  \cdot \hP^{(n)}(\vR) + \sum_{n <  l} \hV_{coul}^{(nl)} 
\end{aligned}
\end{equation}
Note that Hamiltonian \#III is very similar to Hamiltonian \#I in Eq. (\ref{eq:Hsc-Escheme}), the only difference being that we use the transverse E-field $\vE_{\perp}$ in Eq. (\ref{eq:Hsc-Escheme}) instead of $\vE$. For Hamiltonian \# III,  the conserved energy function is defined as $\frac{1}{2}\int d\vR \left(\epsilon_0|\vE|^2+  
\frac{|\vB|^2}{\mu_0} \right)+ \tr{\rho\hH_s^{total}}$, where $\hH_s^{total} = \sum_{n=1}^{N}\hH_s^{(n)} +  \sum_{n <  l}\hV_{Coul}^{(nl)}$.

\begin{sidewaysfigure}
	\centering
	\includegraphics[width=\textwidth]{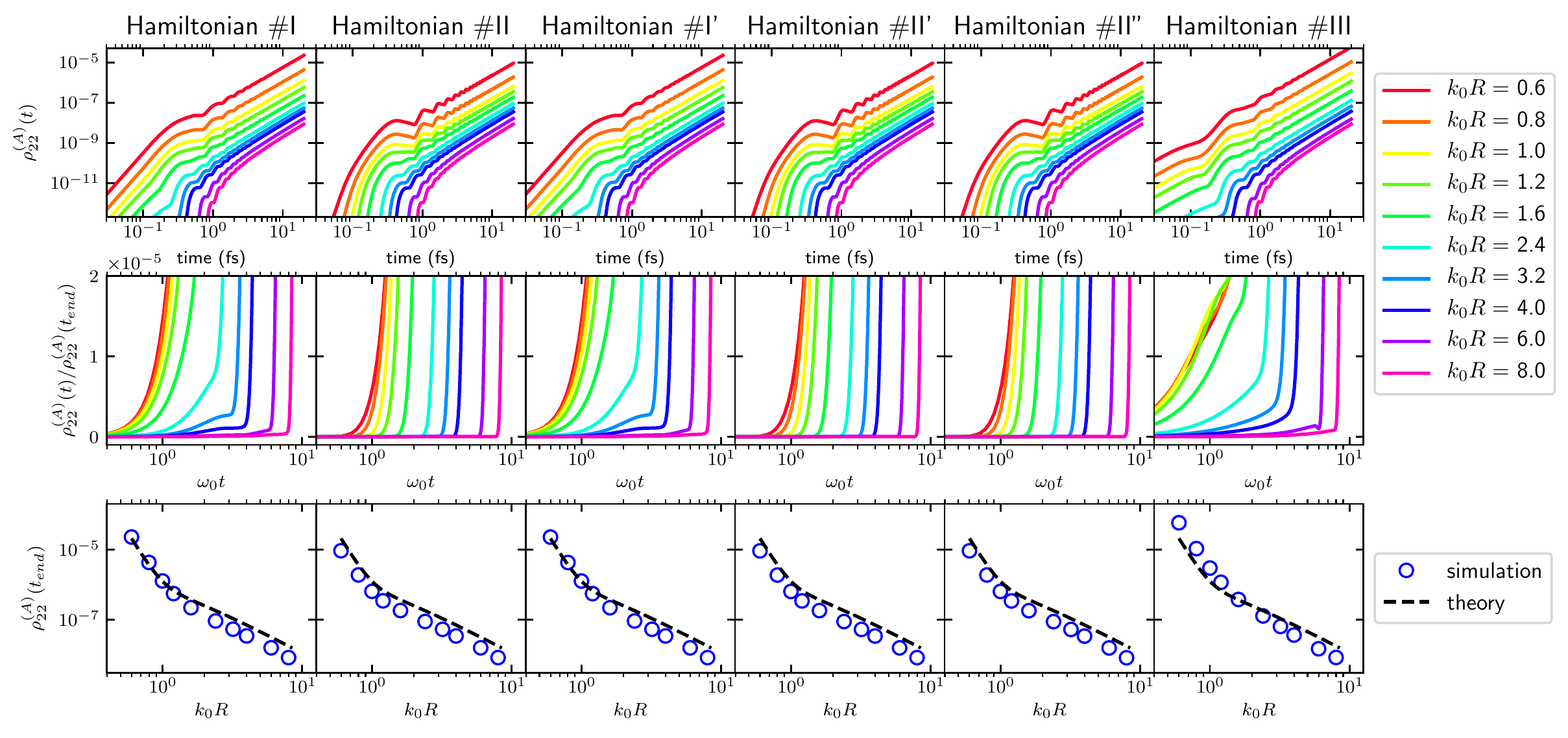}
	\caption{
		Plot of the excited state population of the acceptor ($\rho_{22}^{(A)} (t)$) at short times ($t_{end} = 20$ fs). Results for Hamiltonians \#I, \#II, \#I', \#II', \#II'', \#III are plotted from left to right. All parameters are the same as for Fig. \ref{fig:1}. Note that our results are barely changed if we ignore self-interaction; compare Hamiltonians \#I/II with  Hamiltonians \#I'/II'. However, note that the results for Hamiltonian \#III not only violate causality, but also overestimate the energy transfer at short separation.
	}
	\label{fig:si-1}
\end{sidewaysfigure}

In Fig. \ref{fig:si-1}, we compare  the excited state population of the acceptor ($\rho_{22}^{(A)}(t)$) at short times for Hamiltonians \#I, \#II, \#I', \#II', \#II'' and \#III. We find that the results are barely changed if we ignore the self-interaction terms in Hamiltonians \#I/II. However, if we use Hamiltonian \#III, we find that such an approach not only violates causality, but also leads to an overestimation of energy transfer at short separation. Obviously, because of doubling counting,  Hamiltonian \#III is not recommended.

\begin{figure*}
	\centering
	\includegraphics[width=\textwidth]{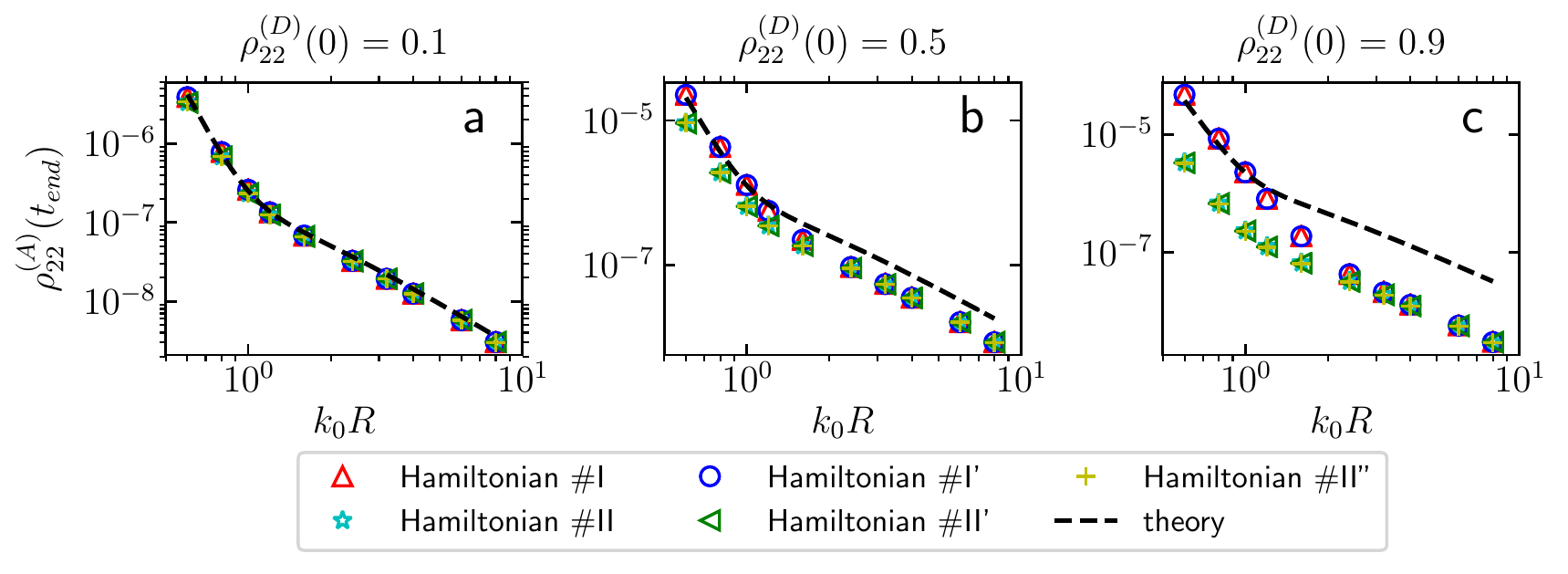}
	\caption{
		Plot of the excited state population of the acceptor at time  $t_{end} = 20$ fs ($\rho_{22}^{(A)}(t_{end})$) versus the intermolecular separation ($k_0 R$) using a  logarithmic scale. Six methods are compared: Hamiltonian \#I (red up triangle), Hamiltonian \#II (cyan star), Hamiltonian \#I' (blue circle), Hamiltonian \#II' (green left triangle), Hamiltonian \#II'' (yellow plus) and QED (Eq. (\ref{eq:P_2_Power}), black dashed line). All parameters are the same as in Fig. \ref{fig:2}. Note that results for Hamiltonians \#I/II are nearly identical to results for Hamiltonians \#I'/II', indicating that self-interaction is not important here.
	}
	\label{fig:si-2}
\end{figure*}

Lastly, in Fig. \ref{fig:si-2}, by changing the initial excitation on the donor in the same manner as in Fig. \ref{fig:2}, we further confirm ignoring self-interaction has a minimal effect on energy transfer.

In Eqs. (\ref{eq:HI_matrix}) and (\ref{eq:HII_matrix}), we give the matrix representations for Hamiltonians \#I and \#II in the basis  \{$\ket{gg}$, $\ket{ge}$, $\ket{eg}$, $\ket{ee}$\}. To facilitate understanding, it is also helpful to show the matrix representations for Hamiltonians \#I' (defined in Eq. (\ref{eq:Hsc_E_noself})), \#II' (defined in Eq. (\ref{eq:Hsc_D_noself})), \#I'' (defined in Eq. (\ref{eq:Hsc-Dscheme_noself})), and \#III (defined in Eq. (\ref{eq:Hsc_EV})) that we discussed in Figs. \ref{fig:si-1}-\ref{fig:si-2}. These matrices  are:
\begin{equation}\label{eq:HI1_matrix}
\hH_{sc}^{I'} = 
\begin{pmatrix}
0 &v_A^{I'} &v_D^{I'} &v \\
v_A^{I'} & \hbar\omega_0 & v & v_D^{I'}\\
v_D^{I'} & v & \hbar\omega_0 & v_A^{I'}\\
v & v_D^{I'} & v_A^{I'} &2\hbar\omega_0
\end{pmatrix}
\end{equation}

\begin{equation}\label{eq:HII1_matrix}
\hH_{sc}^{II'} = 
\begin{pmatrix}
0 &v_A^{II'} &v_D^{II'} & 0 \\
v_A^{II'} & \hbar\omega_0 & 0 & v_D^{II'}\\
v_D^{II'} & 0 & \hbar\omega_0 & v_A^{II'}\\
0 & v_D^{II'} & v_A^{II'} &2\hbar\omega_0
\end{pmatrix}
\end{equation}

\begin{equation}\label{eq:HII2_matrix}
\hH_{sc}^{II''} = 
\begin{pmatrix}
0 &v_A^{II''} &v_D^{II''} & 0 \\
v_A^{II''} & \hbar\omega_0 & 0 & v_D^{II''}\\
v_D^{II''} & 0 & \hbar\omega_0 & v_A^{II''}\\
0 & v_D^{II''} & v_A^{II''} &2\hbar\omega_0
\end{pmatrix}
\end{equation}

\begin{equation}\label{eq:HIII_matrix}
\hH_{sc}^{III} = 
\begin{pmatrix}
0 &v_A^{III} &v_D^{III} & v \\
v_A^{III} & \hbar\omega_0 & v & v_D^{III}\\
v_D^{III} & v & \hbar\omega_0 & v_A^{III}\\
v & v_D^{III} & v_A^{III} &2\hbar\omega_0
\end{pmatrix}
\end{equation}
where $v = \frac{1}{\epsilon_0}\int d\vR \ \vxi^{(D)}_{\parallel}\cdot\vxi^{(A)}_{\parallel}$, $v_{D}^{I'} = -\int d\vR \ \left(\vE + 2\text{Re}\rho_{12}^{(A)}\vxi^{(A)}_{\parallel}\right)\cdot\vxi^{(D)}$,
$v_{D}^{II'} = -\int d\vR \ \vE \cdot\vxi^{(D)}$,
$v_{D}^{II''} = -\int d\vR \ \vD \cdot\vxi^{(D)}$, and
$v_{D}^{III} = v_{D}^{II'}$. $v_{A}^{I'}$, $v_{A}^{II'}$, $v_{A}^{II''}$ and $v_{A}^{III}$ are defined analogously.


\section{Details on deriving Eq. (\ref{eq:P_2_Power}) and Weisskopf-Wigner theory}
If we ignore the factor of $P_2^{(D)}(0)$, Eq. (\ref{eq:P_2_Power}) is derived   from  Eq. (4.7.19) in Ref. \citenum{salam2010molecular}. Now in Ref. \citenum{salam2010molecular}, the initial donor state is chosen as  the excited state $(0, 1)$, so $P_2^{(D)}(0) = 1$. For our purposes, with a donor in a superposition state, we will make the rotating wave approximation so that our final energy transfer expression (Eq. (\ref{eq:P_2_Power})) can be derived by multiplying Eq. (4.7.19) in Ref. \citenum{salam2010molecular} by a factor of  $P_2^{(D)}(0)$.

Note that for solving problems of energy transfer as mediated by EM fields, one standard approach today would be to use Weisskopf-Wigner theory\cite{weisskopf1930calculation, Meystre2007,thakkar2015quantum}, which successfully models light-matter interactions starting with a vacuum photon-field and assuming weak light-matter interactions. To capture RET within such a context, one would need to allow multiple scattering events, which goes beyond the treatment in Ref. \citenum{thakkar2015quantum}.

\section{Regarding causality in Fig. \ref{fig:1}}
Note that, when the pair of TLSs are relatively close ($k_0R \approx 1$), causality is clearly violated by Hamiltonian \#I according to Fig. \ref{fig:1}\textit{c}. That being said, in this regime, one usually ignores all retardation effects. Such a simplification usually makes sense because, if a resonant EM field arrives, the pair of TLSs will feel effectively the same external perturbation (in phase) and all emission between the pair of TLSs will interface constructively or disconstructively, so that the time delay between the two systems is usually not important. In contrast, for the problem studied here, we have imagined that we can prepare the two quantum subsystems asymmetrically, with one initially excited and the other relaxed to its ground state. For such a case, retardation (albeit very small) might indeed be measurable. 

\section{Necessity of averaging Ehrenfest trajectories}
In the letter, we mentioned that a single Ehrenfest trajectory cannot predict the correct time correlation function for the electric field, $\avg{\hE(0)\hE(t)}$.  
For this reason, in the context of nuclear-electronic trajectories, recent nonadiabatic dynamics work (e.g., symmetrical quasi-classical (SQC) method) has focused on averaging Ehrenfest trajectories over an ensemble of different initial conditions, and in many cases (especially with harmonic baths), the results have been encouraging.\cite{Cotton2013,Cotton2014}  When applying SQC to electrodynamics\cite{Li2018Spontaneous}, treating the EM fields as classical operators, results so far have been mostly (but not entirely) encouraging. See Ref. \citenum{Li2018Spontaneous}.

\bibliographystyle{achemso}

\begin{mcitethebibliography}{44}
\bibitem[foo()]{footnoteProve}
	A straightforward proof of Eq. (\ref{eq:Vdd_interaction_point}) is given here.
	Starting from Eq. (\ref{eq:Vdd_exp_1}), $\hV_{Coul}^{(nl)} =
	\frac{1}{\epsilon_0}\int d\vR\ \hP^{(n)}_{\parallel}(\vR)\cdot
	\hP^{(l)}_{\parallel}(\vR) = - \frac{1}{\epsilon_0}\int d\vR\
	\hP^{(n)}_{\perp}(\vR)\cdot \hP^{(l)}_{\perp}(\vR) = -
	\frac{1}{\epsilon_0}\int d\vR\ \hP^{(n)}_{\perp}(\vR)\cdot \hP^{(l)}(\vR)$.
	Applying the definition of the transverse $\delta$-function,
	$\hV_{Coul}^{(nl)} = - \frac{1}{\epsilon_0}\iint d\vR d\vR' \sum\limits_{ij}
	\vmu_i \vmu_j \delta(\vR-\vR^{(n)})\delta_{\perp ij}(\vR - \vR')\delta(\vR' -
	\vR^{(l)}) = - \frac{1}{\epsilon_0}\iint d\vR d\vR' \sum\limits_{ij} \vmu_i
	\vmu_j \delta(\vR-\vR^{(n)})\left[ \frac{2}{3}\delta_{ij}\delta(\vR-\vR') +
	\frac{1}{4\pi |\vR - \vR'|^3}\left(\frac{3\left(\vR_i
		-\vR'_i\right)\left(\vR_j - \vR'_j\right)}{|\vR - \vR'|^2} -
	\delta_{ij}\right)\right]\delta(\vR' - \vR^{(l)})$. Eq.
	(\ref{eq:Vdd_interaction_point}) is derived by evaluating the above double
	integrals.\relax
	\mciteBstWouldAddEndPunctfalse
	\mciteSetBstMidEndSepPunct{\mcitedefaultmidpunct}
	{}{\mcitedefaultseppunct}\relax
	\EndOfBibitem
\bibitem[foo()]{footnoteNew}
	This factor of 2 can be easily explained in the context of Eq. (\ref{eq:kEh}),
	given that we start with initial population 0.5 in the ground state for the
	donor\relax
	\mciteBstWouldAddEndPuncttrue
	\mciteSetBstMidEndSepPunct{\mcitedefaultmidpunct}
	{\mcitedefaultendpunct}{\mcitedefaultseppunct}\relax
	\EndOfBibitem
\end{mcitethebibliography}
\providecommand{\latin}[1]{#1}
\makeatletter
\providecommand{\doi}
{\begingroup\let\do\@makeother\dospecials
	\catcode`\{=1 \catcode`\}=2 \doi@aux}
\providecommand{\doi@aux}[1]{\endgroup\texttt{#1}}
\makeatother
\providecommand*\mcitethebibliography{\thebibliography}
\csname @ifundefined\endcsname{endmcitethebibliography}
{\let\endmcitethebibliography\endthebibliography}{}

\end{document}